\documentclass[prl,aps,twocolumn,preprintnumbers,superscriptaddress]{revtex4}
\usepackage{graphicx}
\usepackage{amsmath}
\usepackage[usenames]{color}
\usepackage{amssymb}
\newcommand{\be}{\begin{equation}}
\newcommand{\ee}{\end{equation}}
\newcommand{\bea}{\begin{eqnarray}}
\newcommand{\eea}{\end{eqnarray}}
\newcommand{\pr}{\partial}
\newcommand{\nno}{\nonumber}
\newcommand{\bse}{\begin{subequations}}
\newcommand{\ese}{\end{subequations}}

\begin{document}
\title{Kinetic Gravity Braiding and axion inflation}
\author{Debaprasad Maity
\footnote{debu.imsc@gmail.com}}
\affiliation{Department of Physics and Center for
Theoretical Sciences, National Taiwan
University, Taipei 10617, Taiwan}
\affiliation{Leung Center for Cosmology and Particle Astrophysics\\
National Taiwan University, Taipei 106, Taiwan}

\begin{abstract}
We constructed a new class of inflationary model with the higher derivative 
axion field which obeys constant shift symmetry. 
In the usual axion (natural) inflation, the axion decay
constant is predicted to be in the super-Planckian regime which
is believed to be incompatible with an effective field theory
framework. With a novel mechanism originating from
a higher derivative kinetic gravity braiding (KGB) of an axion field 
we found that there exist a huge parameter regime in our model 
where axion decay constant could be naturally sub-Planckian. 
Thanks to the KGB which effectively reduces the 
Planck constant. This effectively reduced Planck scale 
provides us the mechanism of further lowering down the speed of an axion
field rolling down its potential without introducing super-Planckian
axion decay constant. We also find that with 
that wide range of parameter values, our model induces almost scale
invariant power spectrum as observed in CMB experiments.

\end{abstract}

\maketitle

 Inflation is an exponential expansion phase of our universe 
in its very early stage of evolution. Even though this 
is by far the only successful mechanism to solve several problems
in stand Big-Bang model of the universe, we still do not have
a fundamental theory which leads to such a mechanism. 
In order to realise such an exponential expansion,
often a scalar field is invoked with an unnaturally
flat potential which has already been proved to be very difficult to construct
in the quantum field theory framework. However It has 
been well accepted that shift symmetry plays a very crucial 
role in the inflationary dynamics. 
It is this symmetry which keeps the potential sufficiently 
flat to realise inflation. 
In this respect usual standard model of particle physics could 
still be a natural framework to study inflation.
A pseudo scalar field called axion may play an important role in 
this regard. This is a hypothetical field associated with
the Pecci-Quinn symmetry which has been introduce to
solve the strong CP problem in QCD in standard model of particle physics. 
This axion field obeys shift symmetry.
By using this axionic shift symmetry
a "natural" inflation had been proposed in\cite{freese}. In spite of 
its viability, observation suggests that axion decay constant 
should be $f \geq 3 M_p$. Question has been raised
on this large $f$ in the effective field theory framework \cite{banks}
and also quantum gravity effect may also ruin the axion symmetry 
at that scale \cite{linde}. Subsequently 
various generalization of the above natural inflation 
scenario has been made \cite{yamaguchi}. Very recently 
some viable phenomenological extension of this natural inflation 
with the sub-Planckian axion decay constant have also been proposed 
\cite{hans,germani} which leads to the resurgence of interest in this subject.
In this letter we will construct another viable model of 
axion inflation which is relying on the higher derivative kinetic 
gravity braiding. There has been lot of studies based on this 
kind of model in the context of inflation which goes by the name of 
G-inflation \cite{yokoyama}, and also in the context of 
dark energy mode building \cite{deffayet,felice}. A very similar 
approach with a non-minimally coupled UV-protected inflationary 
model of axion field has also been proposed \cite{germani}.

We will very closely follow those construction in this letter.
The essential mechanism which has already been pointed out in \cite{yokoyama}
that the kinetic braiding parameter is playing the role of flattening the 
potential in certain region of parameter space. We will see that 
in that range of parameter space we can make our axion decay constant $f$ to be
sub-Planckian by appropriately choosing another sub-Planckian
scale $s$ associated with the kinetic gravity 
braiding (KGB) of our model.

We start with the following action 
\be
{\cal L} ~=~ \frac {M_p^2}{2}  R - X -  M(\phi) X \Box \phi
- \Lambda^4 \left(1 -\cos \left ( \frac {\phi}{f}\right)\right)
\label{action}
\ee
where
$
X = \frac 1 2 \pr_{\mu} \phi \pr^{\mu} \phi$ and $\Box = \frac 1 {\sqrt{-g}}\pr_{\mu}(
{\sqrt{-g}}\partial^{\mu}) 
$. $f$ is the axion decay constant. ${\Lambda}$ is related to the 
axionic shift symmetry breaking scale. We call the term associated 
with the higher derivative action as KGB following \cite{deffayet}. 
One of the interesting properties of this higher derivative
term is that it does not lead to any unwanted degrees of freedom.

Assuming the usual FRW Metric ansatz 
\be
ds^2 = -dt^2 + a(t)^2 (dx^2 +dy^2 + dz^2),
\ee
we obtain the following Einstein equation for the scale factor $a$ 
\bea
H^2 = - H \dot{\phi}^3 M(\phi)-\frac X 3 + \frac 2 3 X^2 M'(\phi)  + \frac {\Lambda^4}3
 \left(1 -\cos \left ( \frac {\phi}{f}\right)\right) 
\eea
and for the axion field 
\bea
\frac 1 {a^3} \frac d {dt} 
\left[a^3\left(1 - {3 H} {M} \dot{\phi} - 2 M' X \right)\dot{\phi}\right]
& =&  \pr^{\mu} \phi \pr_{\mu}(M' X) \nno\\ 
&-& 
\frac {\Lambda^4}{f} \sin \left ( \frac {\phi}{f}\right) .
\eea
Where, $H = {\dot{a}}/a$ is the Hubble constant.

Following the reference \cite{yokoyama} if we consider slow roll condition, 
the scalar field equation turns out to be 
\bea \label{aeq}
3 H \dot{\phi} \left(1 - 3 M(\phi) H \dot{\phi} \right)+ 
\frac {\Lambda^4}{f} \sin \left ( \frac {\phi}{f}\right) =0
\eea
We assume that the inflation is driven by the KGB such that 
the function $M(\phi)$ satisfies 
$|M(\phi) H \dot{\phi}|\gg 1$. This condition will lead us to
\bea
\tau = \frac {M(\phi) \Lambda^4} f \sin \left ( \frac {\phi}{f}\right) \gg 1.
\eea  
Once above condition is satisfied, the expressions for slow roll parameters
turn out to  be 
\bea
\epsilon &=& \frac {M_p^2} {2 f^2 \sqrt{\tau}} \frac {\sin \left( \frac {\phi}{f}\right)^2}
{\left(1- \cos \left(\frac {\phi}{f}\right) \right)^2} ~;~
\eta = \frac {M_p^2} {2 f^2 \sqrt{\tau}} \frac {\cos \left( \frac {\phi}{f}\right)}
{\left(1- \cos \left(\frac {\phi}{f}\right) \right)}  \nno\\
\alpha &=& \frac{ M_{p}}{36} \frac {M'}{M} \left (\frac {4 \epsilon^2}{\tau}\right)^{\frac
 1 4} ~~~;~~~\beta = M_{p}^2 \frac {M''}{M} 
\left (\frac {4 \epsilon^2}{\tau}\right)^{\frac  1 2}  
\eea
As one can see from the above expressions for the slow roll parameters that
KGB function $M$ flattens the axion potential in term of $\tau$. 
As we will see this particular novel effect of KGB
will help us to lower the axion decay constant $f$ into the 
sub-Planckian regime. The condition eq.(7) with 
$\sin \left( \frac {\phi}{f}\right)$ function also tells us that in order to 
maintain those slow roll condition inflation driven by KGB 
has to happen not very close to the maximum of the potential but 
little away from the maximum such that
$\sin \left(\frac {\phi}{f}\right) \simeq {\cal O}(1)$. We will see in
our subsequent analysis that this is indeed the case.  

Keeping in mind the periodic nature of the potential we will study following
two different choices of braiding functions

\begin{figure}[t!]
\includegraphics[width=3.200in,height=2.20in]{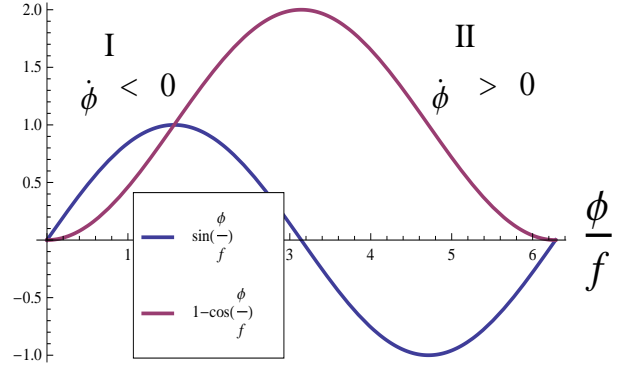}
\caption{\label{fig1} The potential $V(\phi)$ and $V'(\phi)$ up to a 
constant factor related to the amplitude. For the {\bf Model-I}, 
inflation due to KGB occurs in the region I. 
For {\bf Model-II}, 
inflation due to KGB happens on both side of the potential namely 
region (I, II). One can also check that for both the models 
stability condition $M(\phi) \dot{\phi} < 0$ is satisfied. We assume axion 
always rolling down the potential}
\end{figure}

{\bf Model-I}: For $M(\phi) = \frac 1 {s^3}$, where for our subsequent
discussion we fix $s > 0 $ and calling it as our new KGB scale, we get
\bea
\tau_I = \tau_0 \sin \left( \frac {\phi}{f}\right) \gg 1 ~~~;~~~\alpha_I = 0~~~;~~~
\beta_I = 0 ,
\eea
where we define $\tau_0 = \Lambda^4/(s^3 f)$.
This also says that with this particular choice, inflation driven by KGB 
happens in region I of the potential as shown in the above Fig.\ref{fig1}. 
As one can see in this region of the potential speed of axion field 
$\dot{\phi} < 0$. This could further be checked by 
doing perturbation analysis \cite{yokoyama}
that the solution in this region is also stable with the stability condition 
$M \dot{\phi} \simeq \dot{\phi}  <0$.

{\bf Model-II}: On the other hand if we consider $M =  \frac 1 {s^3} 
\sin \left( \frac {\phi}{f}\right)$, we get
\bea \label{kgb}
\tau_{II} = \tau_0 \sin \left( \frac {\phi}{f}\right)^2
 \gg 1 &;& \alpha_{II} = - \frac {M_p}{f} 
\cot \left( \frac {\phi}{f}\right) \left (\frac {4 \epsilon^2}{\tau}
\right)^{\frac
 1 4} \nno \\ \beta_{II} &=& \frac {M_p^2}{36 f^2}
\left (\frac {4 \epsilon^2}{\tau}\right)^{\frac  1 2}
\eea
In this case, on both region (I,II) of the potential (see fig.\ref{fig1})
inflation driven by KGB occurs. In this model also in both regions 
stability condition $M(\phi) \dot{\phi} \simeq \sin 
\left( \frac {\phi}{f}\right) \dot{\phi}  < 0$ is satisfied. 

Similarly, one can choose various possible 
periodic function for $M(\phi)$ which may have different interesting 
phase structure of the inflationary dynamics on various part of 
the axion potential. We will do our detailed study
on those various choices and their dynamics in our future publication. 

This is also interesting to note that for both the models we have
introduced, near the maximum of the potential $\tau$ is very small 
which is essentially referring (see eq.(\ref{aeq},\ref{kgb}))
to the usual slow roll inflationary 
phase of the universe rather than inflation driven by KGB. 
Here we would like to emphasis that if we set our model 
parameter values such that inflationary phase due
to KGB satisfies the CMB observation, then depending 
on the onset value of the axion potential, one can have different types of
inflationary phase along the axion potential with different effect
on the perturbation. One interesting
dynamics would be if onset of inflation happens near the maximum of 
the potential then usual slow roll inflation happens followed by
the inflation driven by KGB. Due to different types of inflationary 
phase along the axion potential it can potentially influence 
the dynamics of various modes of the cosmological perturbation 
in a different manner. 
This can help us to shed some light on the problem of CMB anomaly 
at large angular scales which are associated with 
the long wavelength modes of inflationary perturbation.  
We will defer this study in detail in our future publication. 

If we concentrate on the region of the potential where 
$\sin \left(\frac {\phi}{f}\right)$ being very close to unity, 
then for both the model under consideration, the 
condition of KGB driven axion inflation turns out 
to be 
\bea
\tau_{I} \approx \tau_{II} \approx \tau_0= \frac {\Lambda^4}{s^3 f} \gg 1 \implies s \ll 
\left(\frac {\Lambda^4}{f}\right)^{\frac 1 3}
\eea
So we have enough region of the parameter space where the axion decay constant
could be sub-Planckian along with the KGB scale. In the subsequent
analysis we will derive this scale dependence more rigorously 
taking into account the dynamics of cosmological perturbation.

The region of parameter space could be constrained by the 
dynamics of fluctuation
of the axion field which is directly related to the CMB power spectrum 
$P_{\cal R}$ associated with curvature perturbation ${\cal R}$ and
spectral index $n_s$. The expression for those quantities can be straight 
forwardly calculated as \cite{yokoyama}
\bea
P_{\cal R} =  \frac {3 \sqrt{6}}{64 \pi^2}\frac {H^2}{M_p^2 \epsilon}
~~~;~~~ n_s = 1- 6 \epsilon + 3 \eta + \frac {\alpha} 2
\eea
 
Now let us study the slow roll parameters with respect to both models that
we have discussed above. For both the models the explicit form of the 
spectral index turn out be
\bea \label{spec}
n_s^I &\simeq&  1- \frac {3} {{\cal A}} \frac 
{\sin \left( \frac {\phi}{f}\right)^{\frac 3 2}}
{\left(1- \cos \left(\frac {\phi}{f}\right) \right)^2} 
 + 
\frac {3 } {2 {\cal A}} \frac {\sqrt{\cos \left( \frac {\phi}{f}\right)\cot
\left( \frac {\phi}{f}\right)}}
{\left(1- \cos \left(\frac {\phi}{f}\right) \right)} 
\nno\\
n_s^{II} &\simeq&  1-  \frac {3} {{\cal A}} \frac 
{\sin \left( \frac {\phi}{f}\right)}
{\left(1- \cos \left(\frac {\phi}{f}\right) \right)^2} 
 +  \frac {3} {{2\cal A}} \frac {\cot \left( \frac {\phi}{f}\right)}
{\left(1- \cos \left(\frac {\phi}{f}\right) \right)} \nno\\
&& ~~~~~~~~~~~~~~~~~- \frac 
1 {2\sqrt{{\cal A}}} \cot \left(\frac {\phi} {f}\right) \sqrt{2 \epsilon_{II}}.
\nno
\eea
For our future convenience we have 
further defined a new constant ${\cal A} = \sqrt{\tau_0} (f/M_p)^2$ in the
above expressions. 
One can clearly see from the expression of ${\cal A}$ 
that usual spectral tilt $n_s -1$ of axion (natural) inflation 
is reduced by a factor of $\sqrt{\tau_0}$ or in other words
it essentially suppress the Planck scale. We will see that 
this effectively reduced Planck scale is playing the main roll in 
bringing down the axion decay constant $f$ to be in the sub-Planckian 
regime.

In order to solve the homogeneity and flatness problem of the usual
Big-Bang model, we need to have sufficient amount of inflation. 
This sufficient inflation is measured by so called e-folding number 
${\cal N}= \int_{t_1}^{t_2} H dt$.
From the current cosmological observations the constraint on ${\cal N}
\approx 60$. So further constrain on our model parameters will come from
this e-folding number. The analytic expressions for ${\cal N}$ 
for both the models under consideration turn out to be,
\bea
{\cal N}_I &=& - {\cal A} \left(2 \sqrt{\sin x} + 
\frac {\sqrt {2}\sqrt{1+\sin x}}
{\cos \frac x 2 + \sin \frac x 2} \right. \nno \\ 
&&~~ \left. \times \mbox{EllipticF}
\left[\sin^{-1}\left(\cos \frac x 2 - \sin \frac x 2\right),
 \frac 1 2\right] \right)\Big|_{x_1}^{x_2} \nno\\
{\cal N}_{II} &=&  {\cal A}  \left(x -\sin x \right)\Big|_{x_1}^{x_2} ,
\eea
where we define $x = \phi/f$. In the above expressions, 
the upper limit on the axion field $x_2 = \phi_2/f$ will come from the 
slow roll parameter. As one can imagine that inflation ends when
the slow roll parameter $\epsilon =1 $ which provides us the upper limit.
Furthermore if we set ${\cal N} \approx 60$, 
we can constrain the parameter space of $({\cal A}, \frac {\phi_1}{f})$.
This in turn will constrain the value of the spectral index.
In the above Table-\ref{tab1} we provide some possible 
numerical values of ${\cal A}$ for which we found the values of 
$(\phi_1/f, \phi_2/f, n_s)$ for both the models. As one can see that
the values so obtained for the spectral index $n_s$ are close to the 
observed value from WMAP.
\begin{table}[t!]
\begin{tabular}{|c|c|c|c|c|c|c|} 
\hline
${\cal A}$ & $x_1 =\frac {\phi_1^I}{f}$ & $x_2 =\frac {\phi_2^I}{f} $ 
& $n_s^I $& $ x_1 =\frac {\phi_1^{II}}{f}$ & $x_2 =
\frac {\phi_2^{II}}{f}$ & $n_s^{II}$ \\
\hline
245 & 1.10084 & 0.298338 & 0.970933 & 1.17613 & 0.365871 & 0.98474\\
\hline
215 & 1.15973 & 0.314236& 0.970878 & 1.23119 & 0.382151 & 0.98299\\
\hline
185 & 1.23134 & 0.333563 & 0.970796 & 1.29809 & 0.401779 & 0.98086\\
\hline
125 & 1.43922 & 0.389657 & 0.970436 & 1.49279 & 0.457857 & 0.97474\\
\hline
65 & 1.86299 & 0.504365 & 0.968657 & 1.89973 & 0.569322 & 0.96325\\
\hline
35 & 2.36077 & 0.642477 & 0.961481 & 2.42603 & 0.69966 & 0.94990 \\
\hline
\end{tabular}
\caption{ For both the model under consideration, some specific values
of the parameter ${\cal A}$ which provides us successful inflation driven
by KGB and their possible values of the spectral index.} 
\label{tab1}
\end{table}

Now according to WMAP observations, considering the 
expression for {\bf Model-I}, we know
\bea \label{pr}
P_{\cal R}^I =  \frac {{\cal A}\sqrt{6 }}{32 \pi^2} 
\left( \frac {\Lambda}{M}\right)^4 \frac {(1 - \cos x_1)^3}
{\sin x_1^{\frac 3 2}} \simeq 2.4\times 10^{-9}.
\eea
For a fixed value of ${\cal A}$ the above equation (\ref{pr})
can further provide us a constrain on the value of the axionic symmetry
breaking scale $\Lambda $. As for example if we consider ${\cal A} = 65$,
from the above expression eq.\ref{pr} we found
$\Lambda_I = 5.08 \times 10^{-3}$ in Planck unit for {\bf Model-I}.
With this value of $\Lambda_I$
one can choose one set of values for 
$\{f_I,s_I\} = \{10^{-2},1.634 \times 10^{-5}\}$
in Planck unit such that all the above bounds are satisfied 
with the cosmological observations. A similar estimate 
can be done for {\bf Model-II} where we found 
$\{\Lambda_{II} ,f_{II},s_{II}\} = \{4.99\times10^{-3},
10^{-2}, 1.136\times 10^{-5}\}$ in Planck unit. So one can clearly see that
axion decay constant $f$ as well as the KGB scaling parameter $s$ 
simultaneously could be several order of magnitude lower than the Planck
mass in order to met observational constraints. In addition
another interesting outcome of our construction is that 
we are getting sufficient amount of inflation with the 
value of axion field well below the Planck mass as well. For example
with the above choices of parameters we have $\phi_1^{I} \approx \phi_1^{II}
\approx 1.9 \times 10^{-2}$ in Planck unit.  
Above estimation depends on particular choice of parameters $\{f,s\}$. 
In priciple one has large number of choices
for $\{f,s\}$ as 
\bea \label{ratio}
\frac {s^3}{f} = \frac 1 {{\cal A}^2} \frac {M_p^2}{\Lambda^4}
\eea  

In order to totally fix our model parameters we need to have one more
observable quantity which has non-trivial dependence on the axion
decay constant $f$. For this non-gaussianity would be one of the interesting 
observables which we defer for our future study.
However with the above derived constraint in what follows we will 
discuss about another cosmological observable quantity related to tensor 
perturbation. As one can see from the equation below, 
once we fix the value of ${\cal A}$ 
and scalar spectral index $n_s$, it also fixes the tensor spectral index $n_T$ 
as well as tensor-to-scalar ratio as follows
\cite{yokoyama}
\bea 
n_T &=& - \frac {1} {{\cal A}} \frac
{\sin \left( \frac {\phi}{f}\right)^{\frac 3 2}}
{\left(1- \cos \left(\frac {\phi}{f}\right) \right)^2} \nno\\
r &=&  - \frac {32 \sqrt{6}}{9}\{ n_T^I,n_T^{II}\} = \{0.0757,0.0773\}
\eea
So, the value of tensor-to-scalar ration $r$ for
both the models are very small to be detectable in near future.
We have also checked that as we increase the value of ${\cal A}$, $r$
also increases. Important point to note that perturbation in the tensor
sector does not provide us further constraint on our model parameters. 
We, therefore, are left with one free parameter which can 
probably be fixed if we go beyond the linear cosmological perturbation 
theory.

 In this letter we have discussed a new model of axion inflation which includes 
a specific form of higher derivative terms 
in consistent with the shift symmetry. 
Our model is strongly motivated by the recent studies 
on galileon scalar field theory first introduced in \cite{galileon}.
The specific form of the higher derivative term called   
kinetic gravity braiding is playing the 
crucial role in our model. Interesting point to note that
this particular form of higher derivative term has a
property that it does not introduce any ghost which 
generally appears in a higher derivative theory.
We have seen that this particular form of higher derivative
term helps us to construct a successful axion inflation model with
sub-Planckian axion decay constant. One of the main problem in a 
standard axion (natural) inflation
model is that the axion decay constant $f$ turned out 
to be above the Planck scale in order met CMB observations. 
Through out our current analysis we have shown that this problems
can be easily circumvented by introducing a higher derivative so called KGB 
term in the action for an axion field. This particular KGB term is playing 
the role in pushing the axion decay scale $f$ into the 
sub-Planckian regime. The physical reason behind this mechanism is
coming from the fact that KGB parameter
effectively reduces the Planck constant which in turn makes the speed of
the rolling axion field along its potential slower. 
According to our model we also find a huge parameter region 
where inflation driven by KGB occurs with almost scale 
invariant power spectrum which has already been observed in the 
WMAP experiment. 

We also would like to stress upon the fact that in the linear regime
of cosmological perturbation theory, we could not constrain all our model
parameters. What we infer from our analysis is that once we fix 
the value of our combined parameter ${\cal A}$ from the observed spectral 
index $n_s \geq 0.964$ and axion shift symmetry breaking scale
$\Lambda$ from the observed amount of primordial fluctuation 
$\sim \sqrt{P_{\cal R}} \approx 10^{-5}$, we can fix only the ratio
${s^3}/{f}$ from eq.\ref{ratio} of two scales.
So we can clearly see a huge range of 
values for $\{f,s\}$ where both are sub-Planckian. Interestingly enough 
we obtain sufficient inflation with the axion field value 
well below the Planck constant. These are our main result
in this letter. In order to further constrain the parameters we need to go 
beyond the linear regime like non-gaussianity would be one of such
effects. We defer this analysis for our future study. Another 
interesting effect that could potentially constrain our model is
reheating after the inflation. During our study we overlooked a reference 
\cite{ohashi}, in which a similar approach has been considered. It has
as also been pointed out, that this kind of higher derivative model are
severely constrained by reheating after the inflation,
which may ruin our conclusion.
One of our concerns is that the conclusion arrived
in \cite{ohashi} is based on a particular choice of KGB function.
It would be interesting to check for the other choices such as
the function we have studied in the current paper. 
Detailed study of reheating is beyond the scope of our current paper.
There also exists several other possible mechanisms of reheating 
which are worth studying 
in this scenario if this model is severely constrained by itself from
reheating. Two of the scenarios are curvaton \cite{mukhanov} and modulated 
reheating \cite{kofman}. In both scenarios, there exists additional
light degrees of freedom in addition to the inflaton, which play important
role in the reheating process after the inflation. We keep this for our 
future studies in detail.

{\bf Acknowledgement}\\
I particularly thank our "Sting Cosmology Group" members for various 
stimulating discussion related to this subject.

\end{document}